\documentclass[conference]{IEEEtran}
\IEEEoverridecommandlockouts
\usepackage{cite}
\usepackage{amsmath,amssymb,amsfonts}
\usepackage{algorithmic}
\usepackage{graphicx}
\usepackage{textcomp}
\usepackage{xcolor}
\def\BibTeX{{\rm B\kern-.05em{\sc i\kern-.025em b}\kern-.08em
    T\kern-.1667em\lower.7ex\hbox{E}\kern-.125emX}}
\usepackage[hidelinks]{hyperref}

\begin{document}
\title{Optimized utilization of decentral flexibility for the operational management of cellular multi-modal distribution grids
\thanks{Funding: This research was funded by the I\textsuperscript{3}-Programme of the Hamburg University of Technology (TUHH) in the project `CyEntEE' and the Federal Ministry for Economic Affairs and Climate Action in the project `EffiziEntEE' under the project number 03EI1050A.}
}


\author{
    \IEEEauthorblockN{
        Béla Wiegel\textsuperscript{*},
        Lando Helmrich von Elgott\textsuperscript{**},
        Davood Babazadeh\textsuperscript{*} and
        Christian Becker\textsuperscript{*}
    }
    \IEEEauthorblockA{
        \textsuperscript{*}Institute of Electric Power and Energy Technology, Hamburg University of Technology, Hamburg, Germany \\
        \{bela.wiegel, davood.babazadeh, c.becker\}@tuhh.de
    }
    \IEEEauthorblockA{
        \textsuperscript{**}Energielösungen, Stadtwerke Lübeck Energie GmbH, Lübeck, Germany \\
        lando.helmrichvonelgott@swhl.de
    }
}

\maketitle

\begin{abstract}
Future energy systems rely mostly on supply-dependent resources like wind and solar energy. Since most of the produced energy from distributed renewable sources arises as electricity, electrification of energy consumers and producers, called prosumers, expands. In this context, multi-modal distribution grids make use of the storage capabilities of other energy sectors like heat, and abolish the supply-dependency and uncertainty of the feed-in of decentralized generators. This provides flexibility in the energy system, which increases security of supply and at the same time ensures economical energy provision. The subject of this paper is to exploit the flexibility of such a multi-modal energy system for operation in an optimal way. The cellular approach serves here as the energy system architecture. Aim is to provide flexibility with a cell by optimally distributing a flexibility request to subordinate prosumers changing their active and reactive power. The load change is hereby to be optimally dispatched to the different components of the prosumers under consideration of technical restrictions and dynamic behavior of components, the grids and economical aspects. For dynamic modelling and simulation the open-source modelling language Modelica is used. Using a low-voltage benchmark model expanded by multi-modal technologies, the proposed methodology shows that the cells are capable of providing flexibility to the overlying grid with high accuracy at low costs.
\end{abstract}


\section{Introduction} 
The expansion of sustainable and climate friendly renewable energies in the energy systems worldwide is accelerating significantly. Due to the supply-dependency of renewable resources like wind and solar energy, the integration into the current energy system needs a significant change of operation paradigms~\cite{2020_Torbaghan_Flexibility_Dispatch_Cone_Relaxation}. A large part of the energy arises in a climate-friendly energy system as electricity from distributed generators with low inertia, which at the same time meets increasing electrification of consumers like electrified vehicles for transport and heat pumps for domestic heating \cite{2018_Pieper_Cellular_Cluster_Ancillary_Service, 2022_Sevdari_AS_BEV_Review}. For utilizing of renewable electric energy sources conversion units are of great importance. Hereby, electricity is converted to gaseous energy carriers or enthalpy in form of hot water for heat supply and storages, which leads to the idea of sector coupling resp. multi-modal energy systems. Additionally, the other sectors -- gas and heat -- are capable of supporting the electricity sector during times with poor wind and solar generation and therefore pursue stabilization. Sector-coupled energy systems feature a high variability in possible energy supply paths, and hence, are very flexible.

\begin{figure}
    \centering
    \includegraphics[width=0.8\linewidth]{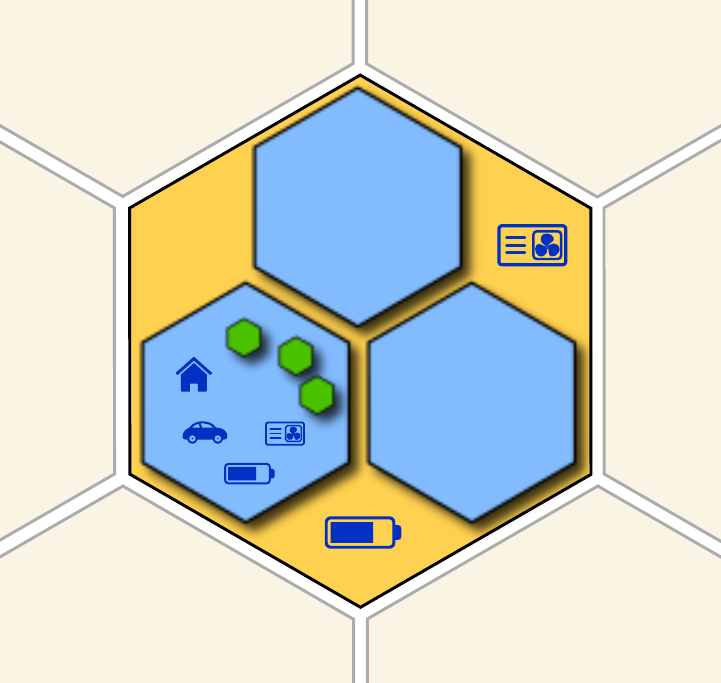}
    \caption{Cellular concept.}
    \label{fig:intro_cellular_concept}
\end{figure}

Sector coupling technologies are relevant on a wide range of size levels, e.g. from Electric Heat Pumps (EHPs), Battery Electric Storages (BESs) and Battery Electric Vehicles (BEVs) on household level, via utility-scale BESs, large EHPs, and combined heat and power on district level, up to large-scale hydrogen storages at transmission grid level. One way for structuring such a multi-modal energy system is the cellular concept\cite{2018_Pieper_Cellular_Cluster_Ancillary_Service, 2019_VDE_Zellularer_Ansatz}. Following this concept, the energy system is structured in two ways: horizontally and vertically. Horizontal cells are created by spatially separating sub-grids which consists of consumers and generators, for example in the form of individual distribution grids below a medium/low-voltage substation of the electricity grid. Physical cells are furthermore coupled functionally via a vertical structure~\cite{2019_VDE_Zellularer_Ansatz}. Hereby, each cell has the ability to fulfill different functions for itself, like self-sufficiency optimization, but also benefit from commissioning tasks to subordinate, adjacent and superimposed cells\cite{2020_Uhlemeyer_Cellular_Approach_Planning}. Fig.~\ref{fig:intro_cellular_concept} visualizes this concept. Providing flexibility to the energy system can be coordinated between the cells, wherewith flexibility provision can be structured~\cite{2019_VDE_Zellularer_Ansatz, 2020_Uhlemeyer_Cellular_Approach_Planning, 2021_Hoth_CyEntEE}.

The spatial-functional structure of the cellular approach allows the distribution of tasks among various grouped generators and loads in order to stabilize the energy system. Even though the cellular approach is not yet common in literature, planning and providing flexibility by distributed generators and loads has been subject of research for a longer period of time. Bartolucci et. al. \cite{2019_Bartolucci_Hybrid_Systems_Ancillary_Services_Households} shows that rule-based and model-predictive control strategies using a linear program for distributed systems on household level are beneficial for grid specific properties like efficiency and voltage control. Sevdari et. al. \cite{2022_Sevdari_AS_BEV_Review} give an overview of needs and possibilities of flexibility provision by BEVs and points out differences in concepts for utilizing BEVs to provide frequency and flexibility services managed by a variety of different stakeholders. In \cite{2021_MansourLakouraj_Microgrid_Flexibility}, MansourLakouraj et. al. formulated a mixed-integer linear program to optimize the costs of micro-grid operation considering ramping rate limitations as a flexibility limit and takes into account the possibility for time-limited uninterruptible switchover to islanding operation.


Taking more technologies into account, Alrumayh et. al. \cite{2019_Alrumayh_Flexibility_Optimization_Residential_Loads} proposes a two-step optimization methodology to optimize energy consumption and flexibility provision on household level together with a grid-related optimal utilization of flexibility to minimize system losses considering power flow equations.

While a variety of different methodologies exists to utilize flexibility of distributed generators and loads, henceforth called prosumer flexibility, just a few consider system dynamics. According to \cite{2021_Degefa_Flexibility_Definition}, there is no relevant literature considering dynamic flexibility, which requires the investigation of needs and technical analysis of flexibility at a time scale below one minute. Due to the increasing dynamics of the electric grid, analysis of the response of generators and loads to time-related changes in power supply or consumption, intrinsic as well as extrinsic, will play an increasing role in future energy systems. In this context, goal of this paper is to distribute a flexibility request in one energy cell to its subordinate cells optimally, thus minimizing operational costs. Limits in ramping-rates of the investigated generators and loads and dynamics of their controllers are considered which is an innovation in this context. This paper answers the following two questions:

\begin{itemize}
    \item How can an energy cell in the low-voltage level consisting of a large number of individual prosumers optimally apply and dispatch a required change of the load resp. generation, considering the dynamics of the prosumers' components?
    \item \begin{minipage}[t]{\linewidth}
    	How can the utilization of flexibility of different prosumers be evaluated?
    \end{minipage}
\end{itemize}

The structure of this paper is organized as follows: section~\ref{sec:flexibility} gives an overview of requirements for provision of flexibility, points out the importance of dynamic energy system modelling and explains the technologies modelled in the scope of this paper. The optimization approach and necessary boundary conditions are described in section~\ref{sec:optimization}, and analyzed on the basis of a use case in section~\ref{sec:use_case}. Final remarks and an outlook are given in section~\ref{sec:conclusion}.

\section{Flexibility in distribution grids considering dynamics} \label{sec:flexibility}

\begin{figure}
    \centering
    \includegraphics[width=1\linewidth]{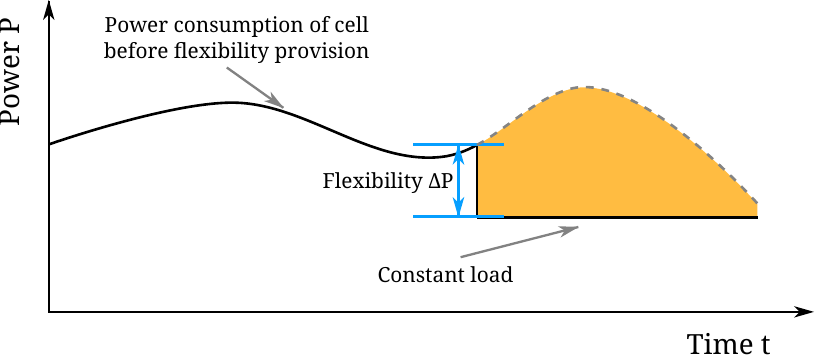}
    \caption{Visualization of the flexibility provision. The yellow marked area between the originally planned power consumption of the cell and the changed power consumption while providing flexibility depicts the overall change in energy consumption.}
    \label{fig:flexibility_simple_time_graph}
\end{figure}

The increasing number of decentralized generators and consumers, called prosumers, connected in the distribution network is growing continuously. To maintain the resilience of the power grid, the technical utilization of the schedulability of the loads of distributed components is an important measure. This is not only due to the supply dependency of generators utilizing renewable primary energy resources, but also due to the situationally high uncertainty of the weather forecast \cite{2022_Moews_Copula}. Thus, the goal of this paper is to change the active and reactive power of an energy cell at the Point of Common Coupling (PCC) on request by a superior cell at a certain point in time, i.e., to provide flexibility by all connected prosumers. The superior cell can also be understood as a system operator. This task is visualized in fig.~\ref{fig:flexibility_simple_time_graph}. Since the real power consumption may vary from the power schedule defined at the market, the power consumption of the cell is to be held constant in this scenario in order to assure a secured power curve to the superimposed grid. The load change is requested by a superior energy cell, which can be considered to be a system operator. Now, a key question to be solved is how to dispatch the flexibility request to the different prosumers within the cell.

\begin{figure}
    \centering
    \includegraphics[width=1\linewidth]{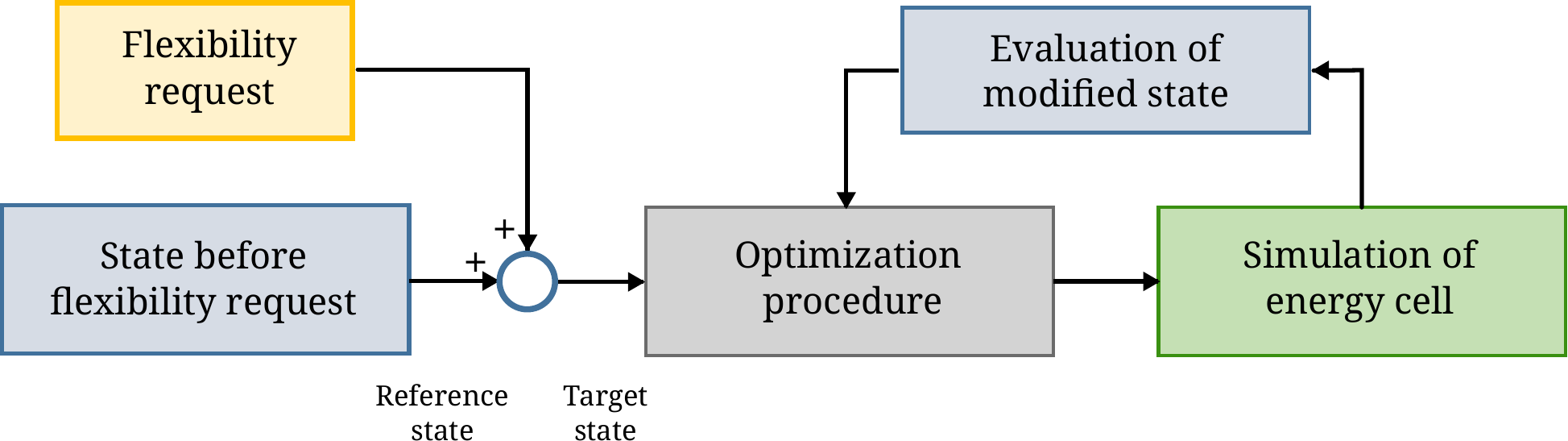}
    \caption{Methodology: Optimized dispatch of a load change by dynamic simulation.}
    \label{fig:flexibility_methodology}
\end{figure}

The provision of flexibility is analyzed using a digital twin of the real system. An energy cell is modeled with technologies most relevant described by a differential-algebraic system of equations, implemented in the acausal modelling language Modelica~\cite{2023_Modelica_Specification, 2021_Senkel_TransiEnt_Status}. Technically, the digital twin is compiled using the Modelica development environment Dymola (Dassault Systèmes) as a stand-alone so called Functional Mock-Up Unit (FMU) connected to a Python interface via the Functional Mock-Up Interface (FMI). Following this approach, it is possible to capture the dynamics of the components in the energy cell, e.g. in the form of limited ramping rates, and to take them into account when providing flexibility. Since this approach considers non-linearities of the system, linear programming is not suitable for optimization. Instead, black-box optimization is chosen. The overall methodology for the optimized dispatch implemented in this paper is depicted in fig.~\ref{fig:flexibility_methodology}.

First, the power consumption of the energy-cell at the PCC is calculated which gives the state of the system before provision of flexibility. This reference state together with the flexibility request results in the target state at the PCC. In the first optimization iteration, the target state at the PCC is given to the optimization procedure for initialization of the power change of the components within the energy cell. After initialization, the new system state while providing flexibility is mapped to the digital twin of the real system. The simulation is started to calculate the system response while providing flexibility. With the system response, the quality of flexibility provision is evaluated for the next optimization iteration. The optimization is done by a black-box optimizer, implemented in SciPy~\cite{2022_SciPY_Basin_Hopping}. The optimizer implements the meta-heuristic Basin Hopping algorithm. The optimization is finished after convergence of the evaluation. In the next section, the parts of the model relevant for the optimization procedure and the optimization procedure itself are explained, before the use case exemplifies how the procedure works.

\section{Optimal flexibility distribution} \label{sec:optimization}
First, basic features of dynamic modeling are explained. Then, the optimization approach is described in more detail.

\subsection{Overview of modelled technologies in Modelica}
    In order to investigate the possibilities to dispatch flexibility on distribution level, a model of a low-voltage distribution grid is implemented in the Modelica modelling language using the open-source TransiEnt Library~\cite{2021_Senkel_TransiEnt_Status}. The bottom-up modelled low-voltage grid consists of a transformer balancing the power demand of the prosumers, which are connected via distribution lines, and the overlying power system. The modelled prosumers are capable of providing  flexibility from multiple sources. In this study private BESs, EHPs an BEVs are considered flexible loads that change their point of operation on request. Furthermore, inverters of PV-systems can fulfill reactive power requests. These flexibility sources are modelled so that they represent the physical behaviour of each technology while also meeting social requirements and ensuring high prosumer comfort.
    
    \subsubsection{Photovoltaics (PV) and Battery Electric Storage (BES)}
        A fraction of the modelled prosumers is generating power, using rooftop solar power systems. Therefore, the generation of power is highly dependent on the weather condition. In this case weather data from Hamburg (Germany) is used~\cite{2021_DWD_ICON_Model}. BESs are used to balance production and consumption and therefore maximize self-use of the generated power on prosumer level as it is economically more feasible. The BESs can also be used as a source of flexibility by adjusting their operation based on flexibility requests. As the solar power generation and BESs are providing DC, an inverter is required for conversion to AC in order to connect to the electrical power lines. The inverters are capable of providing up to 30\% of their power rating in the form of reactive power even when there is no solar generation\cite{2013_SMA_QatNight}. Therefore the inverters of the PV-systems can be used to deliver reactive power flexibility.

    \subsubsection{Electric Heat Pump (EHP)}
        As a rising share of EHPs in heat supply is expected, a fraction of the prosumers is modelled with an air-water-heatpump for domestic heating. The coefficient of performance is calculated based on the Carnot efficiency and a constant effectiveness factor, resulting in a non-linear relation. Each EHP-system is locally controlled at household level to maintain a constant temperature of 45°C in a hot water storage which supplies the households heat demand. For periods of high heat demand or to achieve higher storage temperatures above 50°C an additional direct electric heating element is added to each system. The EHP-systems are capable of providing flexibility by changing the electric power consumption on request resulting in a change of storage temperature. An increase of the electric power consumption will result in excess heat and therefore a gain in the storage temperature up to 90°C. When lowering the electric power consumption the households heat demand is still satisfied by taking heat from the storage causing a decrease in the storage temperature to a minimum temperature of 35°C which the EHP-system will always maintain. Beyond these boundaries, no flexibility can be provided by the EHP-system.
        
    \subsubsection{Battery Electric Vehicle (BEV)}
        Based on the current vehicle fleet in Germany, BEVs are placed in the modelled grid. As the BEVs are used for transportation, they regularly disconnect from the grid and loose battery charge. A BEV is considered disconnected from the grid between the first and the last trip of each day and is unavailable for flexibility purposes during that time.
        When a BEV returns, it reconnects to the grid and starts charging its battery at the rated power of the used charging point. To provide flexibility BEVs are generally capable of reducing their charging power (V1G) while others can even provide vehicle to grid (V2G) services by discharging their battery to supply power to the grid.

    In general, the dynamics of the components are approximated by different combinations of first-order, second-order or delay transfer functions. According to \cite{2021_Lehmal}, this simplification captures the dynamics sufficiently. In Modelica, the first-order transfer function is directly implemented by its differential equation, eq.~\eqref{eq:first_order}:

    \begin{equation} \label{eq:first_order}
        \dot{y} = \dfrac{c\:u - y}{T}
    \end{equation}

    where $c$ is the gain factor, $T$ the time constant, $u$ the set point (input) and $y$ the reaction (output) of the component. The controllers are mostly implemented as continuous PID controllers. The parameterization is done based on literature~\cite{2021_Lehmal} and assumptions.

\subsection{Optimization problem formulation and approach}
    The primary goal of the optimization is to fulfill the flexibility request, e.g. by a system operator, at a certain point in time and secondary to minimize the cost of the flexibility provision. Subject of the optimization is a low-voltage energy cell with a variety of different generation and consumption technologies, named in the section before. The objective function $OF$ is shown in eq.~\eqref{eq:optimization_of},
    
    \begin{align} \label{eq:optimization_of}
        \begin{split}
            OF =
            &\left(\sum_{c \in C} \sum_{i \in n_c} k_c \: \left|\Delta P_{c,i}\right|\right) \\
            &+ k_\mathrm{PCC,P} \: \left| P_\mathrm{PCC}^\mathrm{is} - P_\mathrm{PCC}^\mathrm{target} \right| \\
            &+ k_\mathrm{PCC,Q} \: \left| Q_\mathrm{PCC}^\mathrm{is} - Q_\mathrm{PCC}^\mathrm{target} \right|
        \end{split}
    \end{align}
    
    where $C$ is the set of BES, EHP, BEV (V1G and V2G) and PV inverter, $n_c$ the set of components $c$, $k_c$ the specific cost of a load change for component $c$, $\Delta P_{c,i}$ the difference of the optimized power and reference power, $k_\mathrm{PCC,P}$ resp. $k_\mathrm{PCC,Q}$ penalty costs for not reaching the flexibility request for active resp. reactive power at the PCC of the energy cell and $P_\mathrm{PCC}^\mathrm{is}$ resp. $P_\mathrm{PCC}^\mathrm{target}$ the actual active power resp. optimized power at PCC, and the same for the reactive power $Q$. The first summand of the function represents the costs to be minimized, and the second two summands ensure reaching the flexibility target.

\subsection{Estimation of cost of flexibility}
    The cost of flexibility is still subject of research. In order to choose the price of a specific load change for the following use case investigation the willingness of prosumers to co-create distributed flexibility from different sources is considered. As found in research~\cite{KUBLI2018}, prosumers are willing to provide flexibility from PV batteries and up to a certain battery charge limit also from BEVs. The prosumers' willingness to provide flexibility from distributed heat pumps is the lowest and therefore high monetary compensation is required to obtain this flexibility~\cite{KUBLI2018}. For the following optimization the specific cost of load change are chosen accordingly for every modelled flexible technology to represent this behaviour as shown in table~\ref{tab:optimization_costs}. The lowest compensations are paid for flexibility from inverters as they have no significant impact on the comfort of prosumers. The highest specific cost is related to not meeting the target value at the PCC. This cost can be understood as the market value of flexibility at that time.

    \begin{table}[htbp]
        \caption{Specific cost of load change.}
        \begin{center}
            \begin{tabular}{|l|l|r|}
            \hline
            \multicolumn{1}{|c}{\textbf{Technology}} &
            \multicolumn{1}{|c}{\textbf{Symbol}} &
            \multicolumn{1}{|c|}{\textbf{Value} / 0.1~EUR/kW} \\
            \hline
            BES & $k_\mathrm{BES}$ & $2.78\cdot 10 ^{-4}$ \\ \hline
            Inverter & $k_\mathrm{Inv}$ & $1.38\cdot 10 ^{-4}$ \\ \hline
            EHP & $k_\mathrm{EHP}$ & $7.92\cdot 10 ^{-3}$ \\ \hline
            BEV V1G & $k_\mathrm{BEV,V1G}$ & $5.56\cdot 10 ^{-4}$ \\ \hline
            BEV V2G & $k_\mathrm{BEV,V2G}$ & $9.72\cdot 10 ^{-4}$ \\ \hline
            PCC P & $k_\mathrm{PCC,P}$ & $2.78\cdot 10 ^{-2}$ \\ \hline
            PCC Q & $k_\mathrm{PCC,Q}$ & $2.78\cdot 10 ^{-2}$ \\ \hline
            \end{tabular}
        \label{tab:optimization_costs}
        \end{center}
    \end{table}

\subsection{Solver for the optimization problem}
    Following the approach of dynamic system simulation, the meta-heuristic black box optimization algorithm named \emph{Basin Hopping} for solving the dispatch task is chosen~\cite{1998_Wales_BasinHopping}. The Basin Hopping algorithm is a two-step procedure that is particularly used for solving high-dimensional optimization problems. Basin Hopping is chosen over other meta heuristic optimization techniques since it proofed to give superior results for the optimization task investigated in this study. The optimization is initialized by first estimating one good solution in the solution space. In this case, the reference state serves as the starting point. At the beginning of every Basin Hopping iteration, the algorithm performs a random step through the solution space. Then, the Downhill-Simplex procedure, also called Nelder-Mead method, is started to find the local optimum beginning from this point. After local convergence the local optimum is evaluated and the probability to continue the next iteration from this candidate solution is calculated by the Metropolis Criterion, shown in eq.~\eqref{eq:optimization_metropolis},

    \begin{equation} \label{eq:optimization_metropolis}
        p = \begin{cases}
            1 & \Delta Y \leq 0 \\
            \exp\left(-\dfrac{\Delta Y}{T_\mathrm{BH}}\right) & \Delta Y > 0
        \end{cases}
    \end{equation}

    where $\Delta Y$ is the difference between the new and the previous solution and $T_\mathrm{BH}$ the so called temperature parameter. The higher the dimensionless temperature value gets, the higher the probability for accepting worse solutions. During the optimization process the algorithm is set to aim for an acceptance rate of 50\% by adjusting the step size of the random step between the Basin Hopping iterations. A high value for $T_\mathrm{BH}$ therefore leads to an increase in step size and a strong exploration of the objective function, whereas a low temperature causes a decrease in step size and a strong exploitation.
    
\begin{figure}
        \centering
        \includegraphics[width=\linewidth]{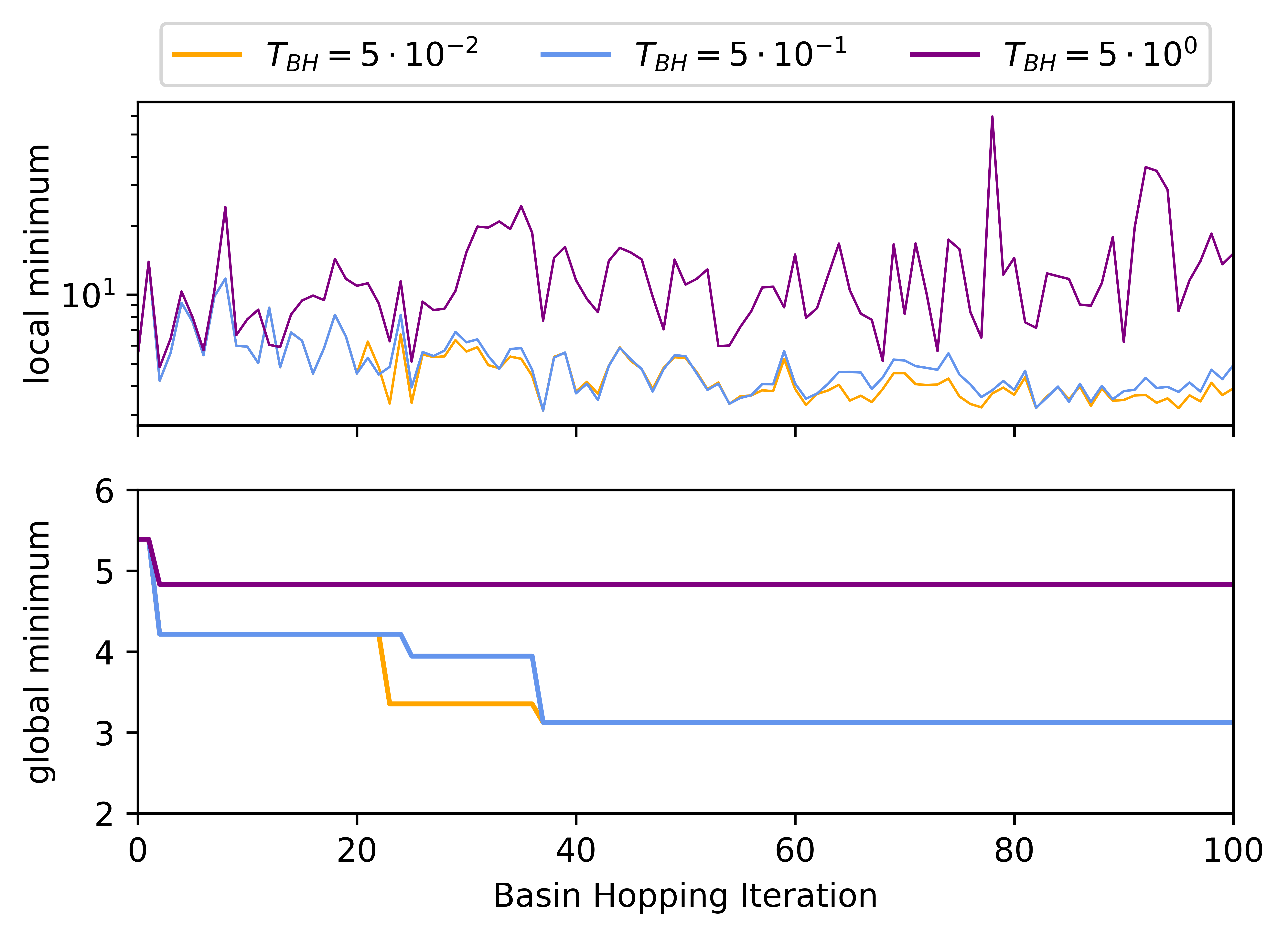}
        \caption{Local an global optimization result along 100 Basin Hopping iterations with different temperature values.}
        \label{fig:convergence_bh}
    \end{figure}

    As the capability of accepting worse solutions during the optimization process is important in order for the algorithm to leave a local minimum, the selection of a suitable temperature value is of great importance for the performance of the Basin Hopping algorithm. In general the magnitude of $T_\mathrm{BH}$ should be comparable to the separation in objective function value between local minima \cite{2022_SciPY_Basin_Hopping}. Since the slope of the objective function is given by the choice of specific cost of load change $k_c$ and penalty costs $k_\mathrm{PCC}$, the temperature parameter has to be chosen accordingly. To do so, the same flexibility request is optimized using different temperature parameters. The results of local an global optimization after every Basing Hopping iteration are shown in Fig~\ref{fig:convergence_bh}. The figure shows the results of the local search in the upper plot while the lower plot shows the lowest yet found optimum which is the result of the global search. It is seen that high values of $T_\mathrm{BH}$ cause the algorithm to enforce exploration resulting in higher values of the local search. In this case better results are found when using low temperature values as this will lead to a strong exploitation. Focusing on exploitation is an advantage when having a good estimated starting solution which is true in this case. Therefore for the following use case investigation a value of $T_{\mathrm{BH}}=0.5$ is chosen. An overview of the most important Basin Hopping parameters is given in table~\ref{tab:optimization_parameters}.

\begin{table}[htbp]
    \caption{Parameterization of Basin Hopping Algorithm}
    \begin{center}
    \begin{tabular}{|l|l|l|}
    \hline
    \multicolumn{1}{|c}{\textbf{Parameter}} &
    \multicolumn{1}{|c}{\textbf{Symbol}} &
    \multicolumn{1}{|c|}{\textbf{Value}} \\
    \hline
    Temperature & $T_\mathrm{BH}$ & 0.5 \\ \hline
    Iterations & $n_\mathrm{iter}$ & 50 \\ \hline
    Step size & -- & adaptive \\ \hline
    Local search & -- &  Nelder-Mead\\ \hline
    Starting value & $x_0$ & power in previous time step \\ \hline
    \end{tabular}
    \label{tab:optimization_parameters}
    \end{center}
\end{table}

\section{Use Case} \label{sec:use_case} 
\subsection{Investigation of Benchmark Grid Model}
\begin{figure}
        \centering
        \includegraphics[width=0.9\linewidth]{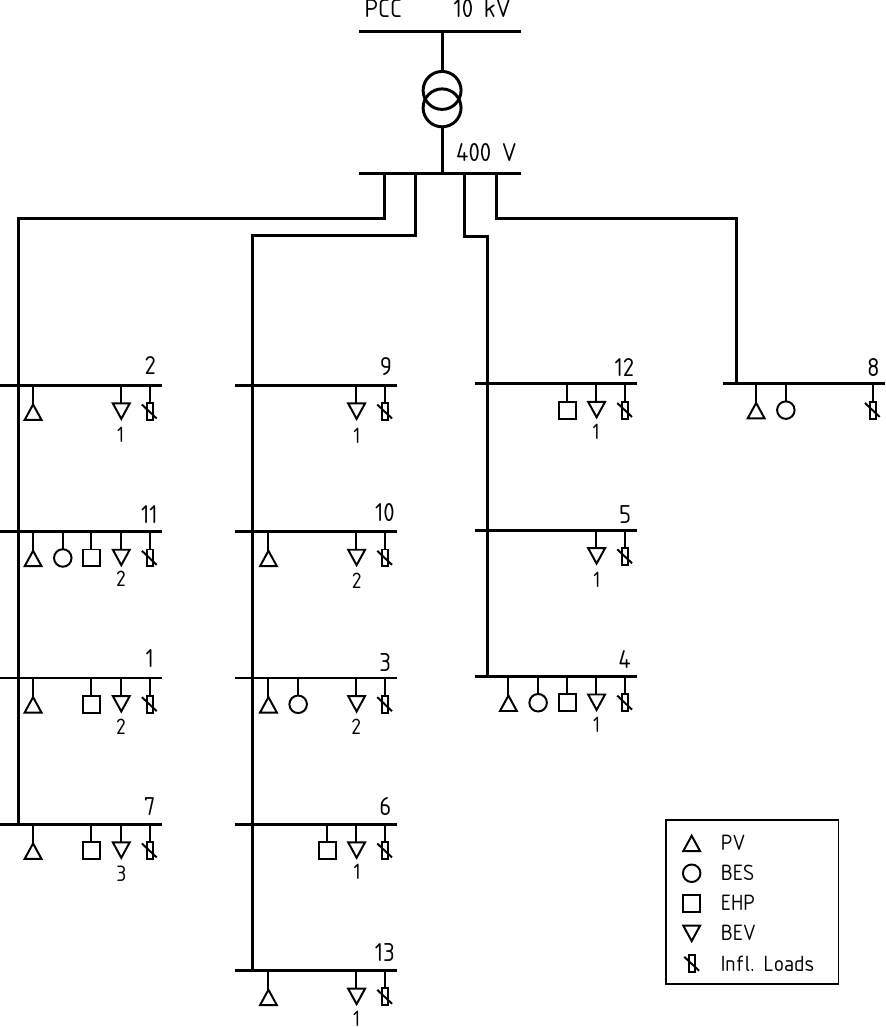}
        \caption{Schematic overview of the investigated grid model. It is based on \textit{rural 1} scenario of SimBench~\cite{2020_SimBench}, expanded by technologies investigated in this paper. The numbered nodes are connected via lines. Each prosumer has different technologies, depicted by symbols. Below BEV, the number of BEVs per prosumer is indicated.}
        \label{fig:use_case_grid_model}
    \end{figure}

    To investigate the potential of using meta-heuristic optimization for flexibility provision on low-voltage level, a model of a low-voltage grid with a large share of prosumers with decentralized renewable energy generation as well as new heating and mobility technologies is built. The grid topology and loads are taken from the rural low-voltage-grid of SimBench, a benchmark library that represents typical grids in Germany \cite{2020_SimBench}. The prosumer loads, taken from SimBench, are considered to be inflexible. The model is then expanded by randomly adding additional technologies for flexibility provision to the prosumers. A schematic of the resulting grid-model is shown in Fig.~\ref{fig:use_case_grid_model}. The modelled grid consists of 13~prosumers of which 10 are operating PV-systems of different sizes. Each of the PV-systems includes an inverter that can provide reactive power. Four of the PV-systems have an additional BES that each prosumer utilizes to balance their own power generation and demand but also can be used to provide active power flexibility. Six prosumers use EHPs with heat storages for domestic heating. A total of 18 EVs and charging points are placed within the grid. Five of the EVs are capable of V2G so that they are not only able to reduce their charging power but even discharge into the grid on request. This results in a total of 38 controllable plants that are distributed in the modelled grid and can be requested to achieve a certain change in power flow. The total change in power flow is evaluated at the PCC which also represents a medium/low-voltage substation.

\subsection{Continuous Optimization of Flexibility Provision}
\begin{figure}
        \centering
        \includegraphics[width=\linewidth]{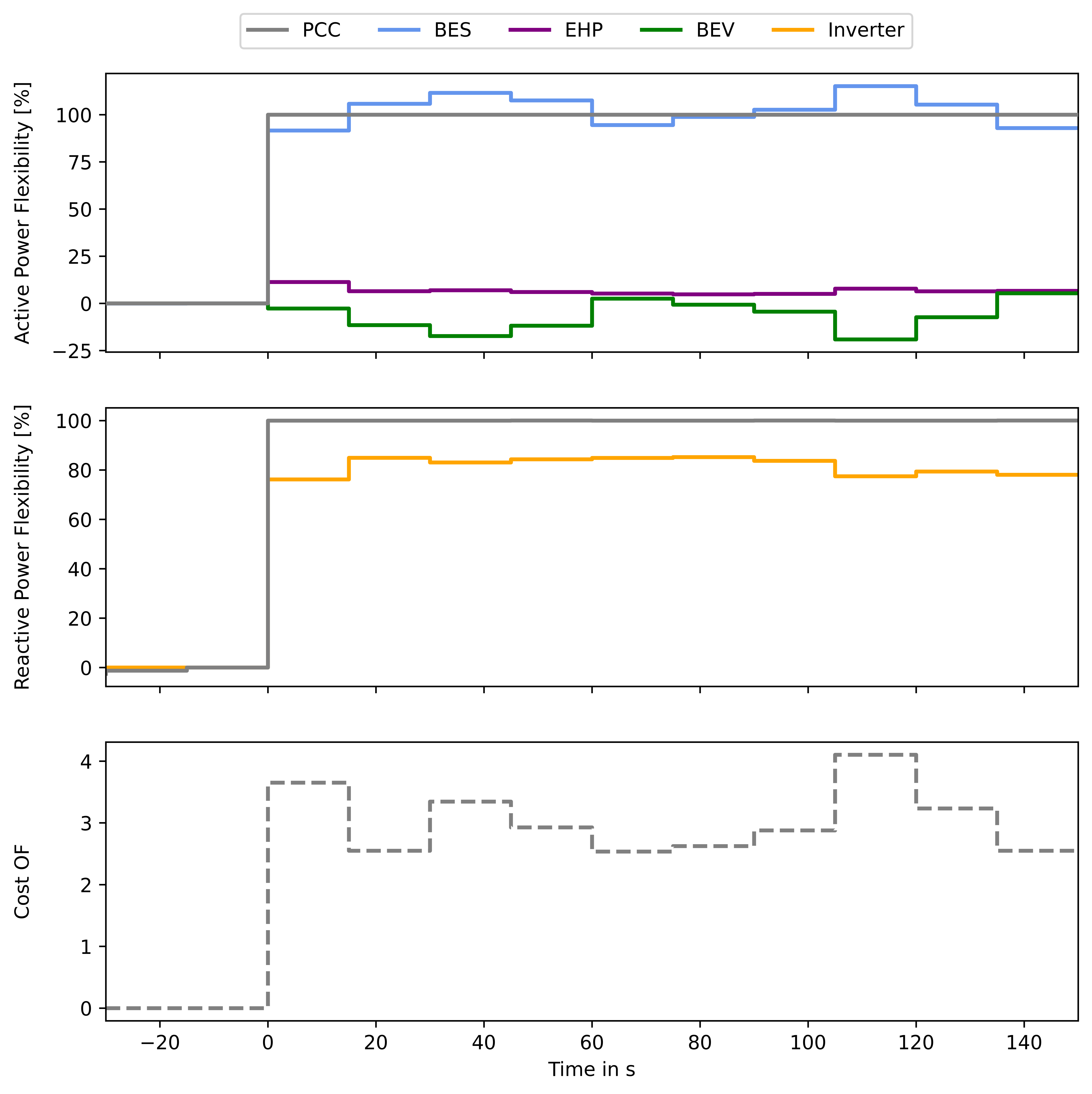}
        \caption{Optimized flexibility response to a request for power gain of $\mathrm{5~kW}$ and $\mathrm{1~kVAr}$ starting at the time $t = 0~\mathrm{s}$. The figure shows the change in power at the PCC and the flexibility provided relative to the total requested flexibility as well as the costs for allocating the flexibility to the prosumers over time.}
        \label{fig:result_Load_Gain}
    \end{figure}

\begin{figure}
        \centering
        \includegraphics[width=\linewidth]{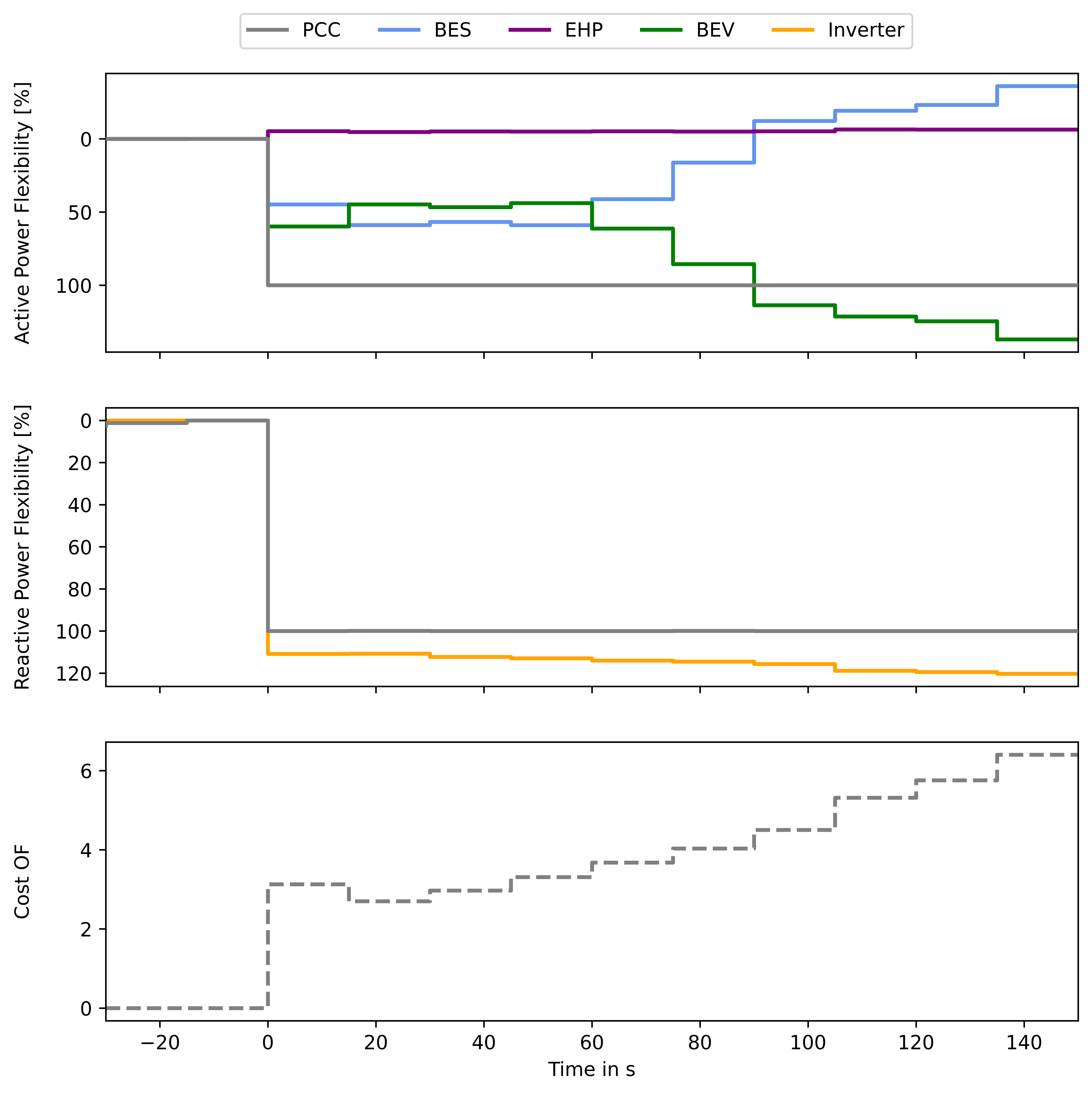}
        \caption{Optimized flexibility response to a request for power reduction of $\mathrm{5~kW}$ and $\mathrm{1~kVAr}$ starting at the time $t = 0~\mathrm{s}$.}
        \label{fig:result_Load_Reduction}
    \end{figure}

    For the use cases, flexibility is requested at a winter evening at 8~pm. The model is simulated seven days ahead of the flexibility request to estimate the conditions of the grid for the use case. The current power flows at all plants and at the PCC right before the flexibility is requested are taken for reference to evaluate the change in power flow. The following time step of 15~seconds is simulated repeatedly each time evaluating the objective function~\eqref{eq:optimization_of}. The dispatch of flexibility to all flexibility options is adjusted iteratively by the Basin Hopping algorithm. The algorithm simultaneously monitors the grid state and only accepts solutions that lie within the physical limitations of the modelled low-voltage lines. When the algorithm finishes optimizing the current time step, it continues with the following time step while still using the original power flows for reference. For the starting point of each time step, the result of the previous time step is used. By this method a continuous distribution of flexibility can be achieved. In Fig.~\ref{fig:result_Load_Gain} and~\ref{fig:result_Load_Reduction} optimization results for different flexibility requests are shown.
    
    Fig.~\ref{fig:result_Load_Gain} shows the optimized response of the modelled grid relatively to the requested load gain of $\mathrm{5~kW}$ and $\mathrm{1~kVAr}$, that equals 100~\%, starting from the time $t = 0~\mathrm{s}$. Before that time no flexibility is provided by the grid. With the first time step of the flexibility provision the provided flexibility of the prosumers causes the desired change in active and reactive power at PCC which is maintained during the entire investigated time interval. The additional active power consumption is mostly provided by the BESs increasing their charging power. The EHPs and BEVs only show minor changes with the BEVs even reducing their charging power at some times opposing the requested flexibility. The change in reactive power at the PCC is provided mostly by the inverters of the PV-systems but not entirely as the reactive power demand at the PCC is also affected by the change in power transmission through the modelled lines and the active power consumption of technologies like the EHPs that work with a considerable power factor. The total costs for providing the requested flexibility only fluctuate slightly as the dispatch of flexibility barely varies during the entire investigated time interval.
    
    Fig.~\ref{fig:result_Load_Reduction} shows the optimization result for a flexible load reduction of $\mathrm{5~kW}$ and $\mathrm{1~kVAr}$. Again, the requested load change in active and reactive power is achieved at the PCC by distributing the flexibility over the flexible prosumer technologies. For this request the BESs capability of providing flexibility is limited as the batteries are barely charged due to poor PV generation. Since this affects the total available power of the BESs the active power flexibility is supplied to the same degree by BESs and BEVs. As the BES is drained entirely during the investigated time interval, the BEVs take over the provision of flexibility and even have to compensate the drained BES as the drainage causes an increase in load in consideration of the reference state. The changed power consumption within the grid also causes the PV-inverters to reduce additional reactive power in order to meet the requested target value at the PCC. The increasing share of flexibility from BEVs and the growing requirement of reactive power supply causes an steady increase of costs during the investigated time interval.

\section{Conclusion and Outlook} \label{sec:conclusion}
Future energy systems will be characterized by a large number of decentralized generators supplying electric power at distribution level. The continuous balancing of load and generation therefore can prospectively not be done centrally by large power plants, but is also the responsibility of many small producers, i.e. prosumers. This requires a high level of regulatory and technical effort to utilize the distributed flexibility of prosumers. A possible approach to address this problem is the cellular architecture, in which the energy system is separated at multiple levels into individual cells. On the one hand, these cells can ensure the balancing of load and generation, and on the other hand they can also provide flexibility for superimposed and adjacent cells.

In this paper, the optimized provision of flexibility by a single cell in the distribution grid is investigated. The aim is to change the load of the cell at a defined point in time. The optimization takes into account both technical criteria, i.e. the fulfillment of flexibility, network restrictions and the condition of the individual components, as well as economic factors such as costs and acceptance for using prosumer flexibility. The modeling is done using a dynamically modeled digital twin of the energy cell based on models of the TransiEnt Library implemented in Modelica. The optimization is done using the meta-heuristic Basin Hopping algorithm which is accessing the model via the FMI standard. The results show that the technical task of flexibility provision can be fulfilled for different use cases. Although it is not possible to objectively examine the quality of the found optimum, the distribution shows expected behavior.

The proposed methodology still has limitations considering the optimization speed and is not yet ready to solve the optimization problem in real-time. The procedure however is capable of optimizing a planned provision of flexibility. Research is currently being carried out on how the numerical efficiency of this methodology can be improved so that the computing time is reduced. The described concept of consecutive optimization steps, wherewith previous optimization results serve as the starting point for following optimization steps, leads already to faster convergence. In addition, it is considered to generate an abstract mathematical model of the energy system model, with which the provision of flexibility is gradually achieved in the context of the temporally continuous real-time application. In such a case, the iterations do not have to be perfect, but only point in the direction of a better solution, conceptually based on the iteration steps of the fast decoupled load flow calculation. Moreover, a research project is currently being planned at TUHH in which the most important components will be investigated in order to further improve the models of the TransiEnt Library and therefore enhance its applications.

\bibliographystyle{IEEEtran}
\bibliography{Paper_Flexibility_Dispatch}

\begin{thebibliography}{10}
\providecommand{\url}[1]{#1}
\csname url@samestyle\endcsname
\providecommand{\newblock}{\relax}
\providecommand{\bibinfo}[2]{#2}
\providecommand{\BIBentrySTDinterwordspacing}{\spaceskip=0pt\relax}
\providecommand{\BIBentryALTinterwordstretchfactor}{4}
\providecommand{\BIBentryALTinterwordspacing}{\spaceskip=\fontdimen2\font plus
\BIBentryALTinterwordstretchfactor\fontdimen3\font minus
  \fontdimen4\font\relax}
\providecommand{\BIBforeignlanguage}[2]{{%
\expandafter\ifx\csname l@#1\endcsname\relax
\typeout{** WARNING: IEEEtran.bst: No hyphenation pattern has been}%
\typeout{** loaded for the language `#1'. Using the pattern for}%
\typeout{** the default language instead.}%
\else
\language=\csname l@#1\endcsname
\fi
#2}}
\providecommand{\BIBdecl}{\relax}
\BIBdecl

\bibitem{2020_Torbaghan_Flexibility_Dispatch_Cone_Relaxation}
S.~S. Torbaghan, G.~Suryanarayana, H.~Hoschle, R.~D{\textquotesingle}hulst,
  F.~Geth, C.~Caerts, and D.~V. Hertem, ``{Optimal Flexibility Dispatch Problem
  Using Second-Order Cone Relaxation of AC Power Flows},'' \emph{{IEEE}
  Transactions on Power Systems}, vol.~35, no.~1, pp. 98--108, jan 2020.

\bibitem{2018_Pieper_Cellular_Cluster_Ancillary_Service}
C.~Pieper, T.~Hess, and J.~Henoch, ``Renewable, cellular energy clusters
  providing ancillary services,'' in \emph{{NEIS 2018; Conference on
  Sustainable Energy Supply and Energy Storage Systems}}, 2018, pp. 1--7.

\bibitem{2022_Sevdari_AS_BEV_Review}
K.~Sevdari, L.~Calearo, P.~B. Andersen, and M.~Marinelli, ``{Ancillary services
  and electric vehicles: An overview from charging clusters and chargers
  technology perspectives},'' \emph{Renewable and Sustainable Energy Reviews},
  vol. 167, p. 112666, oct 2022.

\bibitem{2019_VDE_Zellularer_Ansatz}
{VDE Verband der Elektrotechnik Elektronik Informationstechnik e.V.}, ``{Der
  Zellulare Ansatz: Ein Beitrag zur Konkretisierung des zellularen Ansatzes mit
  Handlungsempfehlungen},'' \emph{VDE Verband der Elektrotechnik Elektronik
  Informationstechnik e.V.}, 2019.

\bibitem{2020_Uhlemeyer_Cellular_Approach_Planning}
B.~Uhlemeyer, J.~Jakob, M.~Zdrallek, C.~Baumann, W.~Well{\ss}ow, J.~Dickert,
  S.~J. Rasti, G.~Blumberg, and A.~Schinke-Nendza, ``{Cellular approach as a
  principle in integrated energy system planning and operation},''
  \emph{{CIRED} - Open Access Proceedings Journal}, vol. 2020, no.~1, pp.
  58--61, jan 2020.

\bibitem{2021_Hoth_CyEntEE}
K.~Hoth, T.~Steffen, B.~Wiegel, A.~Youssfi, D.~Babazadeh, M.~Venzke, C.~Becker,
  K.~Fischer, and V.~Turau, ``{Holistic Simulation Approach for Optimal
  Operation of Smart Integrated Energy Systems under Consideration of
  Resilience, Economics and Sustainability},'' \emph{Infrastructures}, vol.~6,
  no.~11, p. 150, oct 2021.

\bibitem{2019_Bartolucci_Hybrid_Systems_Ancillary_Services_Households}
L.~Bartolucci, S.~Cordiner, V.~Mulone, and J.~L. Rossi, ``{Hybrid renewable
  energy systems for household ancillary services},'' \emph{International
  Journal of Electrical Power \& Energy Systems}, vol. 107, pp. 282--297, may
  2019.

\bibitem{2021_MansourLakouraj_Microgrid_Flexibility}
M.~MansourLakouraj, M.~J. Sanjari, M.~S. Javadi, M.~Shahabi, and J.~P.~S.
  Catalao, ``{Exploitation of Microgrid Flexibility in Distribution System
  Hosting Prosumers},'' \emph{{IEEE} Transactions on Industry Applications},
  vol.~57, no.~4, pp. 4222--4231, jul 2021.

\bibitem{2019_Alrumayh_Flexibility_Optimization_Residential_Loads}
O.~Alrumayh and K.~Bhattacharya, ``{Flexibility of Residential Loads for Demand
  Response Provisions in Smart Grid},'' \emph{{IEEE} Transactions on Smart
  Grid}, vol.~10, no.~6, pp. 6284--6297, nov 2019.

\bibitem{2021_Degefa_Flexibility_Definition}
M.~Z. Degefa, I.~B. Sperstad, and H.~S{\ae}le, ``{Comprehensive classifications
  and characterizations of power system flexibility resources},''
  \emph{Electric Power Systems Research}, vol. 194, p. 107022, may 2021.

\bibitem{2022_Moews_Copula}
S.~Möws, M.~Ahrens, and C.~Becker, ``{Comparing R-Vine Copulas and Quantile
  Regression Forests for Reliability Forecasting of Renewable Energies},'' in
  \emph{2022 IEEE 21st Mediterranean Electrotechnical Conference (MELECON)},
  2022, pp. 872--877.

\bibitem{2023_Modelica_Specification}
\BIBentryALTinterwordspacing
{Modelica Association}. (2023) {Modelica Language for Systems Modeling --
  Language Specification}. [Online]. Available:
  \url{https://specification.modelica.org/master/}
\BIBentrySTDinterwordspacing

\bibitem{2021_Senkel_TransiEnt_Status}
A.~Senkel, C.~Bode, J.-P. Heckel, O.~Sch{\"u}lting, G.~Schmitz, C.~Becker, and
  A.~Kather, ``{Status of the TransiEnt Library: Transient Simulation of
  Complex Integrated Energy Systems},'' in \emph{Proceedings of 14th Modelica
  Conference 2021, Link{\"o}ping, Sweden, September 20-24, 2021}, ser.
  Link{\"o}ping Electronic Conference Proceedings.\hskip 1em plus 0.5em minus
  0.4em\relax {Link{\"o}ping University Electronic Press}, 2021, pp. 187--196.

\bibitem{2022_SciPY_Basin_Hopping}
\BIBentryALTinterwordspacing
SciPy. (2022) {Python Library Version 1.9.3}. [Online]. Available:
  \url{https://docs.scipy.org/doc/scipy/reference/generated/scipy.optimize.basinhopping.html#scipy.optimize.basinhopping}
\BIBentrySTDinterwordspacing

\bibitem{2021_DWD_ICON_Model}
D.~Reinert, F.~Prill, H.~Frank, M.~Denhard, M.~Baldauf, C.~Schraff,
  C.~Gebhardt, C.~Marsigli, and G.~Zängl, ``{DWD Database Reference for the
  Global and Regional ICON and ICON-EPS Forecasting System, Version 2.1.7},''
  Deutscher Wetterdienst, Tech. Rep., 2021.

\bibitem{2013_SMA_QatNight}
\BIBentryALTinterwordspacing
{SMA Solar Technology AG}. (2013) {Profitable Nachtschicht für
  Wechselrichter}. [Online]. Available:
  \url{https://www.sma.de/fileadmin/Partner/SMA_Connect/WP_QATNIGHT_ADE132110W.pdf}
\BIBentrySTDinterwordspacing

\bibitem{2021_Lehmal}
C.~Lehmal, ``{Investigation and Validation of Stability for the Photovoltaic
  Intergration into a Medium Voltage Grid Based on PHIL Testing},'' Master's
  thesis, Graz University of Technology - Institute of Electrical Power
  Systems, 2021.

\bibitem{KUBLI2018}
\BIBentryALTinterwordspacing
M.~Kubli, M.~Loock, and R.~Wüstenhagen, ``{The flexible prosumer: Measuring
  the willingness to co-create distributed flexibility},'' \emph{Energy
  Policy}, vol. 114, pp. 540--548, 2018. [Online]. Available:
  \url{https://www.sciencedirect.com/science/article/pii/S0301421517308704}
\BIBentrySTDinterwordspacing

\bibitem{1998_Wales_BasinHopping}
D.~Wales and J.~Doye, ``{Global Optimization by Basin-Hopping and the Lowest
  Energy Structures of Lennard-Jones Clusters Containing up to 110 Atoms},''
  \emph{The Journal of Physical Chemistry A}, vol. 101, 04 1998.

\bibitem{2020_SimBench}
S.~Meinecke, D.~Sarajli\'c, S.~R. Drauz, A.~Klettke, L.-P. Lauven, C.~Rehtanz,
  A.~Moser, and M.~Braun, ``{SimBench --- A Benchmark Dataset of Electric Power
  Systems to Compare Innovative Solutions based on Power Flow Analysis},''
  \emph{Energies}, vol.~13, no.~12, p. 3290, Jun. 2020.

\end{thebibliography}

\end{document}